\documentclass[aps,twocolumn,showpacs,floatfix]{revtex4-1}
\usepackage{graphics,epsfig,amsfonts,amssymb,amsmath,soul,natbib}
\usepackage{subcaption}
\usepackage[export]{adjustbox}
\begin{document}

\title{Avalanche correlations and stress-strain curves in discrete dislocation plasticity}

\author{Henri Salmenjoki$^1$, Lasse Laurson$^{2}$ and  Mikko J.\  Alava$^{1,3}$}
\affiliation{$^1$Aalto University, Department of Applied Physics, PO Box 11000, 00076 Aalto, Finland\\
$^2$Computational Physics Laboratory, Tampere University, P.O. Box 692, FI-33101 Tampere, Finland\\
$^3$NOMATEN Centre of Excellence, National Centre for Nuclear Research, A. Soltana 7, 05-400  Otwock-Swierk, Poland}

\begin{abstract}
The sequence of deformation bursts during plastic deformation exhibits scale-free features. In addition to the burst or avalanche sizes and the rate of avalanches the process is characterized by correlations in the series which become manifest in the resulting shape of the stress-strain curve. We analyze such features
of plastic deformation with 2D and 3D simulations of discrete dislocation dynamics models and we show, that only with severe plastic deformation the ensuing memory effects become negligible. The role of past deformation history and dislocation pinning by disorder are studied. In general, the correlations have the effect of reducing the scatter of the individual stress-strain curves around the mean one. 

\end{abstract}

\maketitle

\section{Introduction}
The fact that plastic deformation takes place by avalanches or bursts makes for an interesting connection between the theory of avalanches in dislocation systems and materials science~\cite{papanikolaou2017avalanches, alava2014crackling,zaiser2006scale}. 
In the case of crystalline solids, plasticity is mediated by the stress-driven collective dynamics of dislocations, i.e., line-line topological defects of the crystal lattice.
The paradigm of avalanches implies scale-free features that the statistical properties of the bursts - such as their sizes and durations and often the intra-avalanche waiting times - follow. For ensembles of dislocations in plastically deforming crystals, these signatures of critical-like dynamics are often linked either to the system being driven to the proximity of a depinning-like phase transition~\cite{ovaska2015quenched,salmenjoki2020plastic}, or exhibiting glassy dislocation dynamics~\cite{ispanovity2014avalanches,lehtinen2016glassy}.

For a single 
micron-scale sample undergoing deformation the consequence of a series of bursts  
is an irregular stress-strain curve~\cite{uchic2004sample,dimiduk2005size,ispanovity2010submicron,papanikolaou2012quasi}.
Single crystals containing assemblies of discrete dislocations have the property that, both in models~\cite{ispanovity2014avalanches,lehtinen2016glassy,csikor2007dislocation} and in mechanical tests such as nano/micropillar compression~\cite{dimiduk2006scale,uchic2009plasticity}, the plastic deformation bursts have a broad distribution of sizes. The stress-strain curve then consists of a sequence of such bursts, separated by quiescent periods during which the applied stress is increased. The resulting stress-strain curve has a random appearance, and indeed these processes are most often characterized simply by probability density functions of burst sizes and stress increments, a description ignoring possible correlations in the burst sequence~\cite{zaiser2006scale}. 

Here, we go beyond such a simple characterization of fluctuating crystal plasticity, by studying to what extent stresses, stress increments and bursts sizes along the stress-strain curves are correlated. This is inherently coupled to the question how one may exploit the statistics to reconstruct stress-strain curves \cite{kapetanou2015statistical,szabo2015plastic}.  Related questions include how the stress-strain curves compare with the mean one~\cite{ispanovity2013average}, in particular it is interesting to ask if single systems have a tendency to converge to the mean behavior and how this depends on the degree of plastic deformation. This is in turn dependent on how random is random-looking plasticity, highlighting the need to measure correlations in the avalanche activity. To this end, we investigate how the correlations in avalanche activity influence the plastic deformation by means of 2D and 3D discrete dislocation dynamics (DDD) simulations. Our work extends previous studies of avalanche correlations in plasticity~\cite{kapetanou2015statistical,weiss2003three} and in the related problem of interface depinning in disordered media~\cite{le2020correlations}. The main issues we address are four. First, as a function of plastic strain, we quantify the variation of individual samples around the average (stress-strain curve) behaviour. We then measure the correlations in the series of bursts, focusing on subsequent avalanches. These correlations are important but diminish along the stress-strain curve. This results in a tendency to approach the average response. In the initial stages of plastic deformation these correlations leave an imprint on both the yield stress and on its variation from sample to sample. An important question is how these memory effects work in general. To investigate this, we also study the influence of past deformation by "pre-strained" systems and that of the presence of a competing dislocation pinning mechanism by a field of precipitates. This is related to the general question of how to classify memory effects in physical systems \cite{Keim2019}, in particular in systems undergoing deformation \cite{Keim2020,Mungan2019,Pashine2019}. Our results tackle the subtle effects that arise from the exact configuration of dislocations at the beginning of a deformation experiment or trial: what are the consequences, and how does that depend on the past history. 

The structure of this paper follows the usual paths: first the methods (Section~\ref{sec:methods}), then the results (Section~\ref{sec:results}), and finally the conclusions (Section~\ref{sec:conclusions}).

\section{DDD simulations}
\label{sec:methods}
\begin{figure*}
    \centering
    \includegraphics[width=0.4\linewidth]{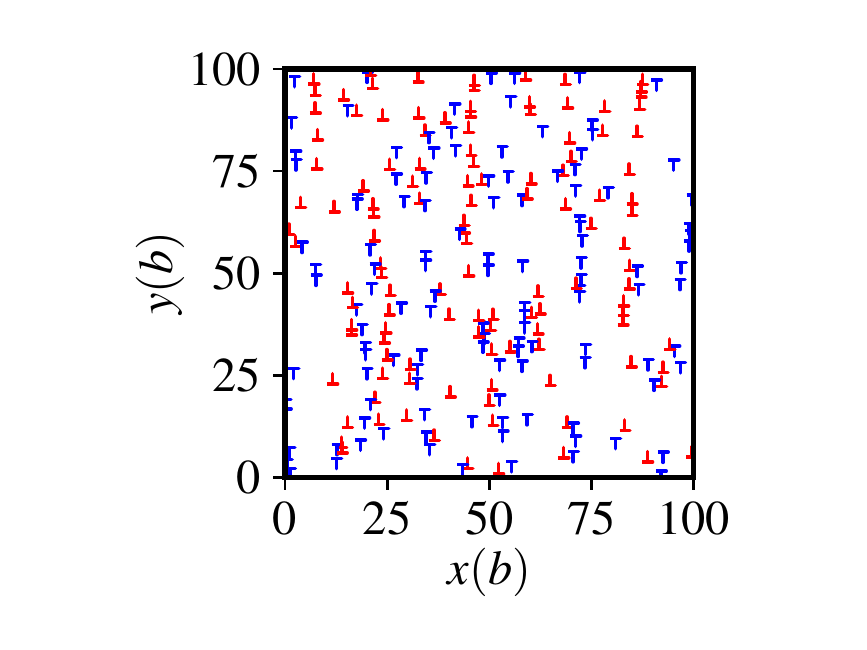}
    \includegraphics[trim={0 0 4cm 0},clip,width=0.4\linewidth]{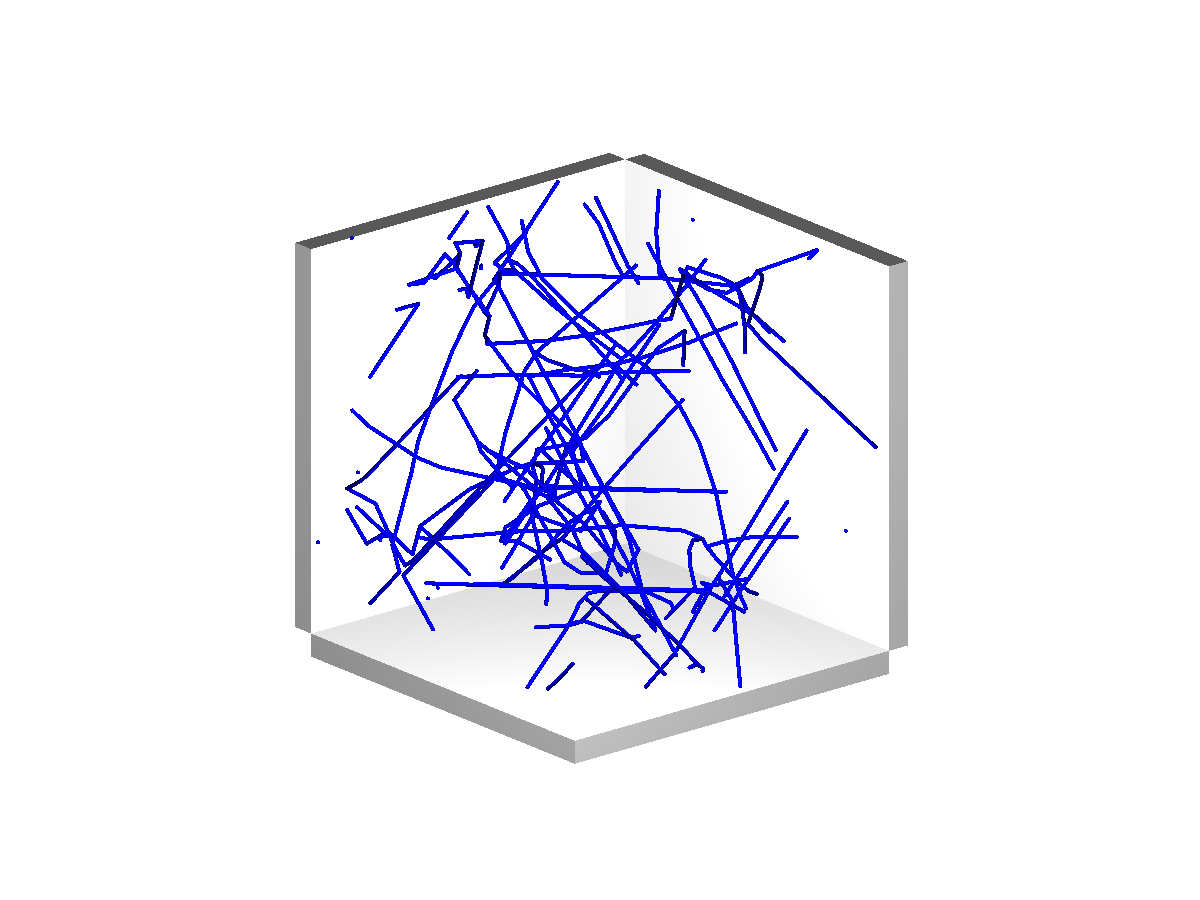}
    \caption{ (a), (b) Examples of initial 2D and 3D  dislocation structures, respectively.  }
    \label{fig:system}
\end{figure*}

To study the correlations between dislocation avalanches, we collect four datasets: We perform standard DDD simulations in both two and three dimensions, pre-strained systems (deformation history) in two dimensions and systems with 
quenched pinning points/precipitates in three dimensions.
For the two-dimensional (2D) systems, we use an in-house developed code to model a square-shaped cross-section of a crystal with infinitely long, parallel edge dislocations. The model is similar to the one studied in several previous works~\cite{ispanovity2014avalanches,ovaska2015quenched,laurson2010dynamical,laurson2012dynamic,salmenjoki2018machine}.
The dislocation are randomly initialized with either positive or negative Burgers vectors with magnitude $b$.
Inside the simulation box, the dislocations are restricted to move in their glide planes in $x$-direction with an equation of motion
\begin{equation}
    \frac{1}{\chi b} v_i = s_i b \left[ \sigma_{ext} + \sum_{i \neq j} s_j \sigma_d(\mathbf{r}_j-\mathbf{r}_i)\right],
\end{equation}
where $s_i$ is the sign of the Burgers vector, $\chi$ is the mobility, $\sigma_{ext}$ is the external stress and the sum is over all other dislocations in the system. 
Moreover, the interaction arising from the shear stress field is given by
\begin{equation}
    \sigma_d(\mathbf{r}) = \frac{\mu b}{2 \pi (1-\nu)} \frac{x (x^2-y^2)}{(x^2+y^2)^2},
\end{equation}
where $\nu$ is the Poisson ratio and $\mu$ is the shear modulus. 
The system is implemented with periodic boundaries and parameters are set to correspond to the simulations with the largest system size found in \cite{salmenjoki2018machine}, i.e. we initialize the system with 400 dislocations in a box with size $L=100 b$ and model parameters are chosen so that $\frac{\mu}{2 \pi (1-\nu)}=b=\chi=1$ and timestep is set to $0.1$.
Therefore, stress and strain are measured in dimensionless units. 
An example of a 2D DDD system is illustrated in Fig. \ref{fig:system}a.

\begin{figure*}
    \centering   
    \includegraphics[width=\linewidth]{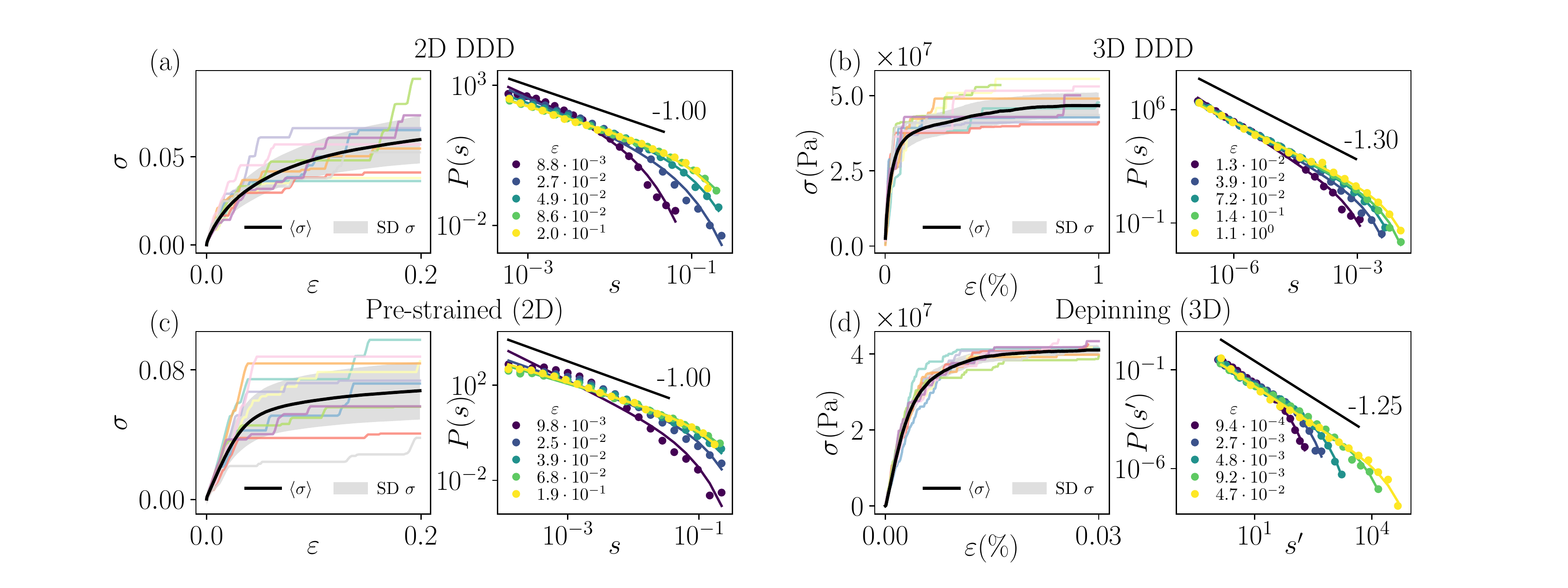}
    \caption{ Stress-strain curves (left) and strain-resolved avalanche size distributions (right)  for (a) 2D (b) 3D (c) Pre-strained 2D and (d) 3D systems with precipitates. The average stress-strain curve (solid black line) is accompanied by single system stress-strain curves (colored lines) and the standard deviation around the average (shaded region). The size distributions for avalanches in different strain bins (color coding) are plotted with the fitting results to Eq. \ref{eq:size_distribution} (solid lines). The solid black lines are added as guide for the eye.  }
    \label{fig:ss_s_distribution}
\end{figure*}

The three-dimensional (3D) DDD simulations are conducted with our version of the ParaDiS code~\cite{arsenlis2007enabling}. 
In ParaDiS, rectangular systems with dislocations are simulated and the program discretizes dislocation lines into a set of nodes and straight segments.
Interaction stresses between the segments are derived from linear elasticity and the long-range nature of these stresses is taken into account by considering periodic images of the system. 
At the dislocation core, interactions are computed with output from MD simulations. 
For the standard 3D DDD simulations (Fig. \ref{fig:system}b), we set the simulation parameters to the values of FCC aluminium ($b=0.286\,\mathrm{nm}$, $\nu=0.35$, $\mu=26\,\mathrm{GPa}$) in a cubic system with size $L_{3D}=1.43\, \mu\mathrm{m}$ and 40 initial straight mixed dislocations \cite{lehtinen2016glassy}.
Unlike 2D DDD, we measure the 3D DDD results in SI-units. 
On the other hand, the disordered systems (i.e. ones with quenched pinning points included~\cite{lehtinen2016multiscale}) are performed along similar lines as in Ref.~\cite{salmenjoki2020plastic}. 
The disorder is implemented as spherical coherent precipitates (with radius $r_p = 28.6\, \mathrm{nm}$) that form  Gaussian barriers for dislocation motion, i.e. the (radial) precipitate-dislocation interaction force is given by
\begin{equation}
    F(r) = - \nabla U(r) = \frac{2Ab^3 r e^{-\frac{r^2}{r_p^2}}} {r_p^2},
\end{equation}
where $A$ is a parameter characterizing precipitate strength. 
The materials parameters used are those of aluminium (like in the standard case), but the system size is $L_{depinning} = 4\, \mu \mathrm{m}$ and initial number of dislocations is 24.
The precipitate parameters are chosen so -- $A = 10^{10}\,\mathrm{Pa}$ and density $\rho_p = 10^{20}\, \mathrm{m}^{-3}$ -- that the system dynamics is dominated by dislocation (de)pinning  \cite{salmenjoki2020plastic}. 

All systems are driven with the quasistatic stress-controlled loading scheme. 
This means that, once we have initialized the systems with randomly placed dislocations (2D) or generated structures using \textit{paradisgen} which is distributed with ParaDiS (3D), and these dislocations have found a meta-stable state after relaxation with $\sigma_{ext}=0$, we start to increase the external stress with a rate $\dot{\sigma}$. 
Only exception is the pre-strained 2D case, where the initial states are prepared by first running a stress ramp until a pre-strain of $\varepsilon_{ID}=0.2$, after which the system is let to relax again at zero stress.
In 2D simulations (both basic and pre-strained), we set $\dot{\sigma}_{2D}=2.5\cdot 10^{-7}$, in 3D $\dot{\sigma}_{3D}=2.5\cdot 10^{13} \mathrm{Pa}/\mathrm{s}$ in the $[010]$ direction, and in 3D with precipitates $\dot{\sigma}_{Depinning}=1.0\cdot 10^{14} \mathrm{Pa}/\mathrm{s}$ in the $[100]$ direction. 
As $\sigma_{ext}$ increases, we measure the velocity of dislocations inside the system (in 2D simply the sum of the $|v_i|$'s, while in 3D we measure the absolute value of the extensive velocity $V(t) = \sum_i l_i v_{\perp, i}$ where $l_i$ is the segment length and $v_{\perp, i}$ velocity perpendicular to the line direction of the segment). 
If the velocity signal then passes a preset threshold $V_0$, an avalanche starts and the stress increments are stopped until the dislocations again reach a jammed state and motion ceases. 
This way the systems produce stress-strain curves with staircase-shape, as can be seen in Fig. \ref{fig:ss_s_distribution} \cite{szabo2015plastic}.
The avalanches are visible as the constant $\sigma$ plateaus and we measure their sizes $s$ by the strain accumulated, i.e. change in $\varepsilon$ (except for the systems with precipitates, for which the avalanche size is the integral of the velocity signal, i.e. $s' = \int_0^T V(t)-V_{0}\, \mathrm{d}t$).

\section{Results}
\label{sec:results}

\subsection{Avalanche analysis}

We then proceed to analyze the avalanches that come in many sizes as seen in Fig. \ref{fig:ss_s_distribution} which also shows the strain-resolved size distributions for the studied systems.
In principle, the systems exhibit three types of avalanches: 
First, there are the small avalanches arising from numerical noise and velocity signal oscillating above and below the avalanche threshold - these are already cropped out from the figures.
Second, there are the power-law avalanches  which, as the name suggests, follow the distribution closely and are mostly uncorrelated. 
Finally, the largest avalanches are the so called cut-off avalanches as there the distributions start waning from the pure power-law.

Because the largest avalanches have also the largest impact on the stress-strain curves, in what follows we restrict our analysis to them. 
However as we search for correlations between avalanches and only few cut-off avalanches occur per system, we take this into account as we define the threshold for the large avalanches.
To define large avalanches systematically, we first fit the (strain-dependent) distributions with the typical form
\begin{equation}
    P(s) = s^{-\tau} e^{-s/s_0},
    \label{eq:size_distribution}
\end{equation}
where both $\tau$ (power-law exponent) and  $s_0$ (the cutoff avalanche size) are fitting parameters. 
Then we set the threshold of large avalanches to $c \cdot s_0$ with $c$ some constant smaller than unity. 
Finally as we have obtained values for $s_0$ in the different strain bins, we can interpolate $c \cdot s_0(\varepsilon)$ to obtain an estimate if any specific avalanche belongs to the large avalanches depending on its starting strain. 

\begin{figure*}
    \centering
    \includegraphics[width=0.32\linewidth]{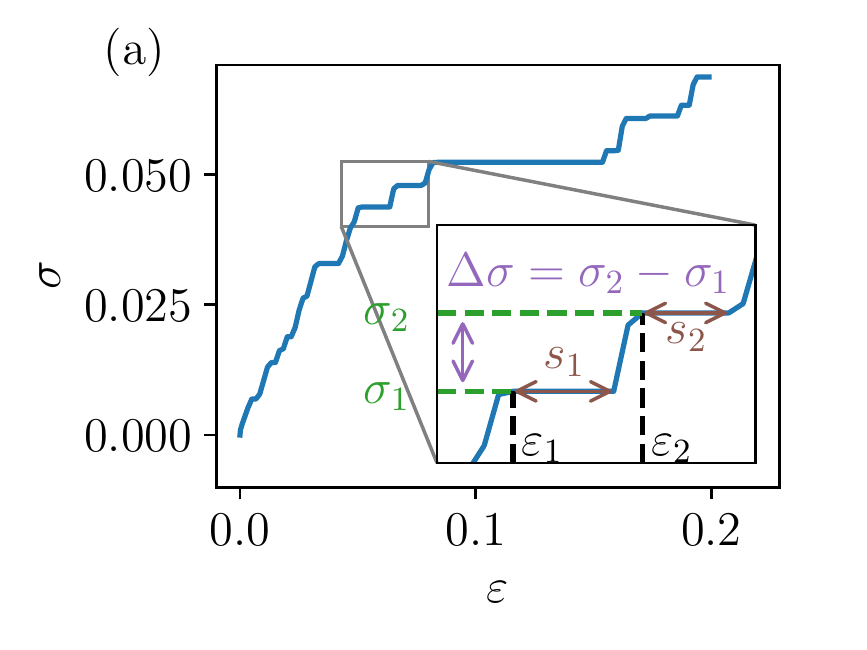}
    \includegraphics[width=0.32\linewidth]{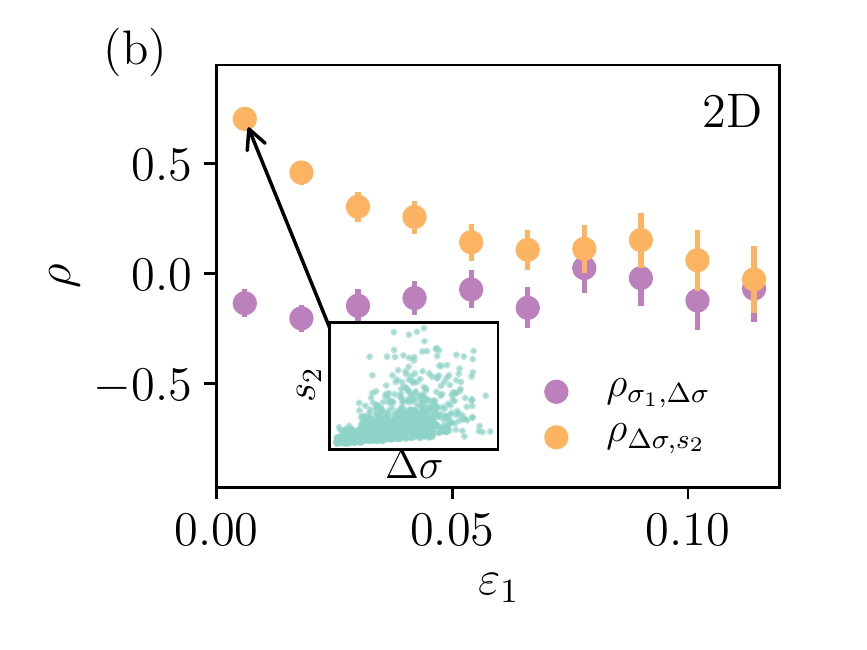}
    \includegraphics[width=0.32\linewidth]{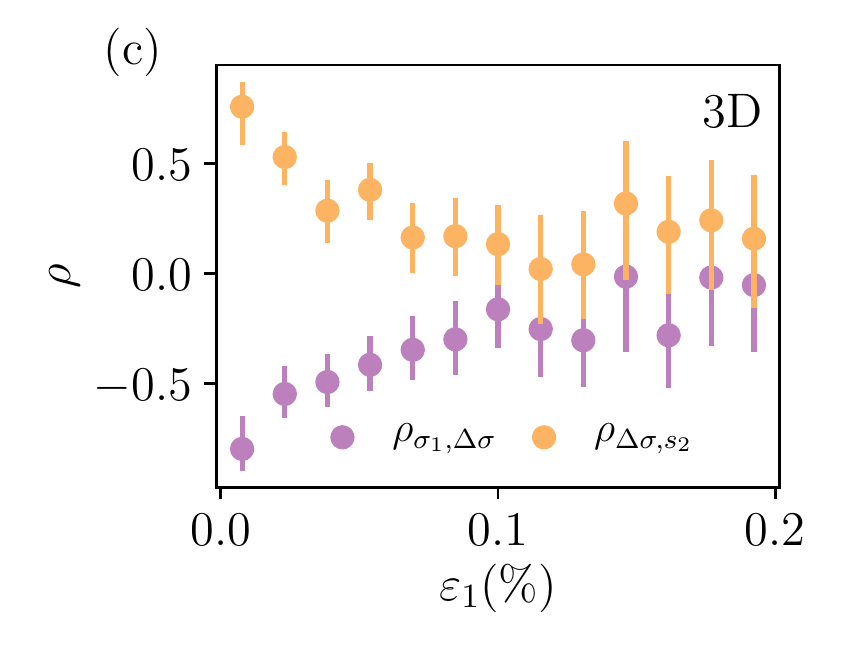}
    \caption{(a) Illustration of the parameters of the two subsequent avalanches. (b), (c) Spearman correlation coefficient between $(\sigma_1, \Delta \sigma)$ and $(\Delta \sigma, s_2)$ as a function of the first avalanche starting strain for avalanches from 2D and 3D systems, respectively.  }
    \label{fig:correlation}
\end{figure*}

From the obtained set of large avalanches, we collect subsequent avalanches that occur in the same system to study correlations between them.
We focus on parameters introduced in Fig. \ref{fig:correlation}a, namely the starting strains of the preceding and following avalanches, $\varepsilon_1$ and $\varepsilon_2$, the sizes of the preceding and following avalanches $s_1$ and $s_2$, and the stress increment between the avalanches $\Delta \sigma = \sigma_2 - \sigma_1$. 
To measure the correlations, we use the Spearman rank correlation coefficient $\rho$. 
Unlike the more commonly used Pearson correlation coefficient, which measures the linear relationship between the values of two variables, Spearman correlation coefficient considers the rank of the variables, thus resulting in +1 (or -1) if the variables form a perfect monotonously increasing (or decreasing) curve, and 0 if no correlation between the ranks exists \cite{spearman1904proof}.
For the case of correlations between (power-law distributed) avalanches, assuming linear dependencies  seems unnecessarily restrictive and, thus, Spearman correlation coefficient is preferred here. 

\subsection{Avalanche correlations in 2D and 3D DDD}

We continue to study the correlations between two subsequent large avalanches. 
The observations made here come with avalanches collected from 5000 2D DDD and 1000 3D DDD (from which 250 have been driven to $\varepsilon \sim 1\%$ and the rest to much smaller strains) systems. 
In Figs.~\ref{fig:correlation}b-c we have the Spearman correlation coefficient for the 2D and 3D cases, respectively, when $\rho$ are computed between the starting stress of the previous avalanche $\sigma_1$ and the stress increment $\Delta \sigma$, and between the stress increment and the size of the following avalanche $s_2$. 
Moreover, $\rho_{\sigma_1, \Delta\sigma}$ and $\rho_{\Delta \sigma, s_2}$ are plotted as a function of the starting strain of the preceding avalanche, $\varepsilon_1$, to see how the correlations change along the stress-strain curve.

Starting from $\rho_{\Delta \sigma, s_2}$, we see that with small $\varepsilon_1$ in both 2D and 3D there is a strong monotonously increasing dependence between the variables as $\rho>0.5$ in both cases (for reference, the inset of Fig.~\ref{fig:correlation}b shows the scatter plot of the variables). 
As $\varepsilon_1$ increases and the systems approach the plateau of the stress-strain response, the correlations disappear.
Then with $\rho_{\sigma_1, \Delta\sigma}$ there is an opposite, monotonously decreasing dependence between $\sigma_1$ and $\Delta \sigma$ which similarly weakens with strain, although at small strains the correlation is much weaker in 2D ($\sim -0.2$ at best) than in 3D ($<-0.6$).  
The correlations measured here are obtained with large avalanche thresholds $c=0.25$ and $c=0.1$ for 2D and 3D, respectively,
and we have tested that the results seem robust with respect to the value of $c$ used or the number of strain bins. 
Other avalanche parameter pairs (e.g. $s_1$ and $s_2$) show no similar, notable correlations.

\begin{figure*}
    \centering
    \includegraphics[width=0.4\linewidth]{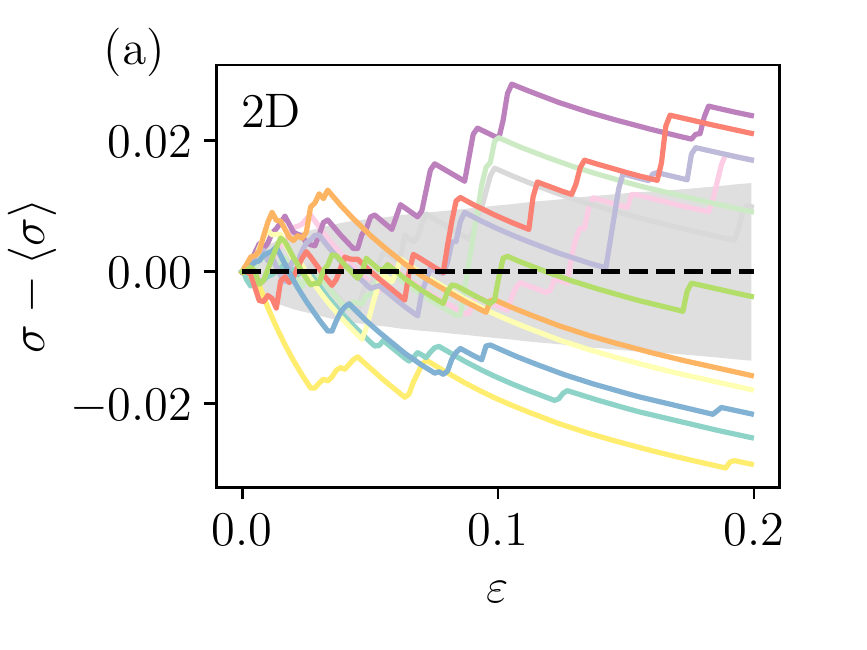}
    \includegraphics[width=0.4\linewidth]{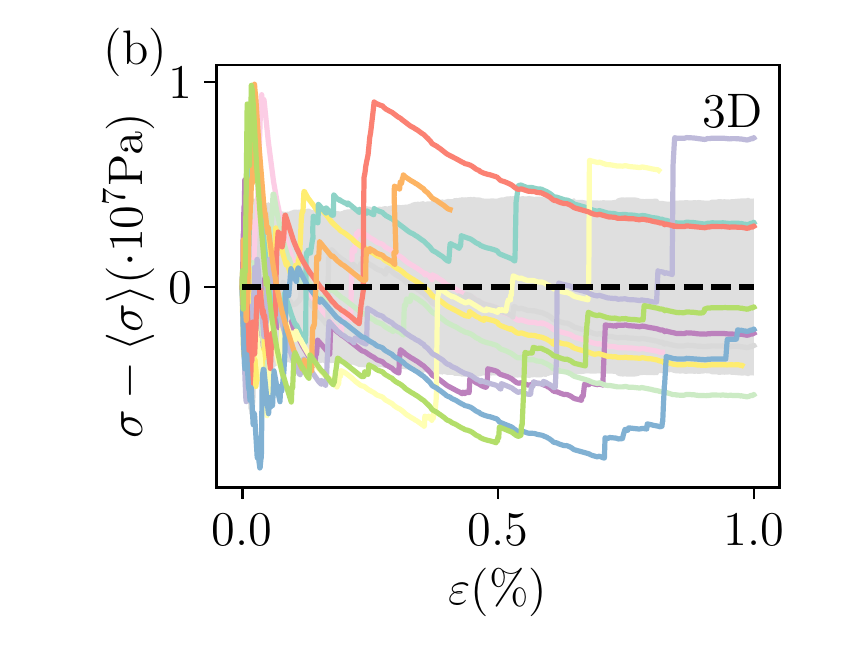}
    \caption{A randomly chosen set of stress-strain curves shifted with the average curve, for 2D (a) and 3D (b) systems. The shaded region corresponds to the standard deviation.   }
    \label{fig:ss_shifted}
\end{figure*}

The correlations have a significant impact on the systems' stress-strain response with small strains.
This is because both $\rho_{\sigma_1, \Delta\sigma}$ and $\rho_{\Delta \sigma, s_2}$ have the same effect of pushing the stress-strain curve towards the system average:
If the preceding avalanche starts with a larger stress $\sigma_1$, then there is a smaller increment in stress before the next avalanche and vice versa, due to the negative $\rho_{\sigma_1, \Delta\sigma}$.
Similarly if the stress increment between the avalanches is large (small), next avalanche will be large (small) as  $\rho_{\Delta \sigma, s_2}$ is positive. 
The push towards the average curve is visualized in Figs. \ref{fig:ss_shifted}a-b which show randomly chosen stress-strain curves of single 2D and 3D systems shifted by the average $\sigma(\varepsilon)$. 
In the figure, intersections of the single curves with the average ($\sigma - \langle \sigma \rangle =0$) are most frequent with small strains where the correlations are stronger, and cease as the correlations vanish. 

\begin{figure*}[h]
    \centering
    \includegraphics[width=0.4\linewidth]{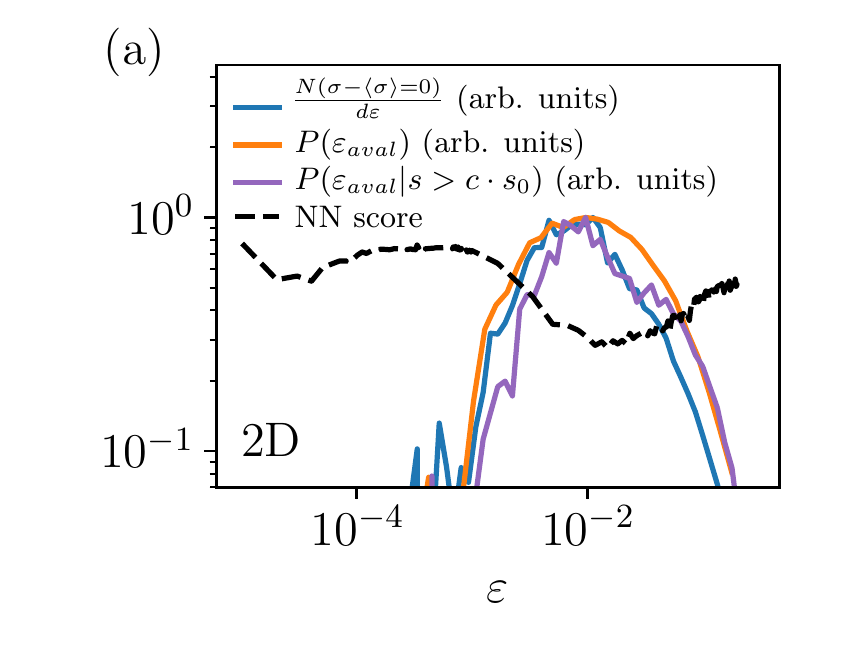}
    \includegraphics[width=0.4\linewidth]{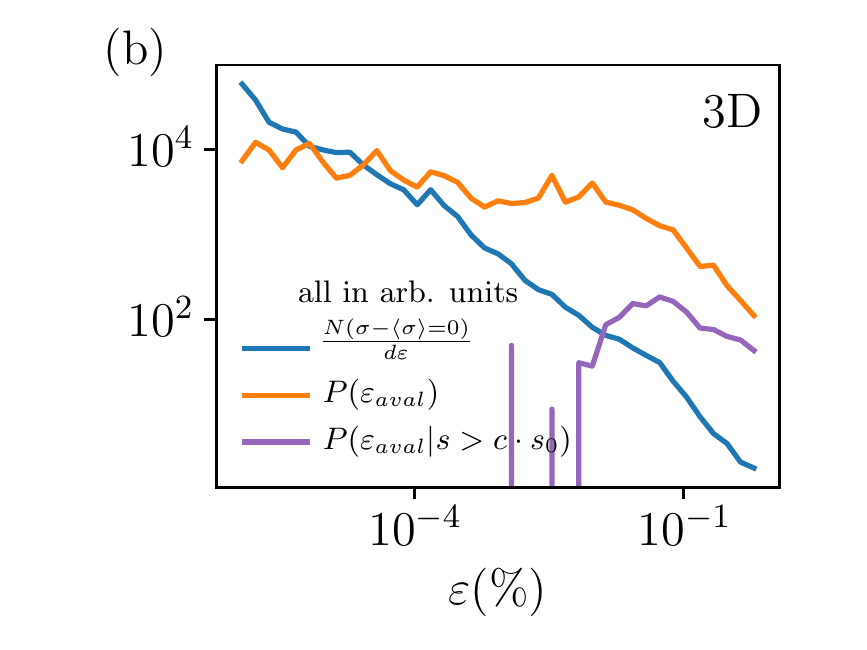}
    \caption{The rate of intersections with the average stress-strain curves $\frac{N(\sigma - \langle \sigma \rangle = 0)}{\mathrm{d}\varepsilon}$ and the distribution of avalanche starting strains in (a) 2D and (b) 3D systems. For 2D systems, the dashed line also shows the goodness of a neural network (NN) prediction of the stress-strain curve as a function of strain as obtained in \cite{salmenjoki2018machine}.  }
    \label{fig:intersectionrate}
\end{figure*}

To elaborate, Fig. \ref{fig:intersectionrate} shows the average rate $\frac{N(\sigma - \langle \sigma \rangle = 0)}{\mathrm{d}\varepsilon}$ along with avalanche starting strain distributions.
Indeed the rate decreases with large strains and it follows $P(\varepsilon_{aval})$ in both 2D and 3D.
But there are some fundamental differences between the two sets.
In 2D, the rate of intersections and avalanche activity is confined to a section of strains, while in 3D the rate and avalanche activity are in an ongoing decrease. 
This difference in the small strain behaviour is mostly explained by considering the simulation protocols: 
the 2D system are relaxed before the loading to a state where all the motion stops in the limit of numerical error. 
However the 3D systems reach no such state as there remain some oscillation and slow decay of dislocation motion, thus resulting in (small) bursts already in the start of the stress ramp. 

\begin{figure*}[h]
    \centering
    \includegraphics[width=0.4\linewidth]{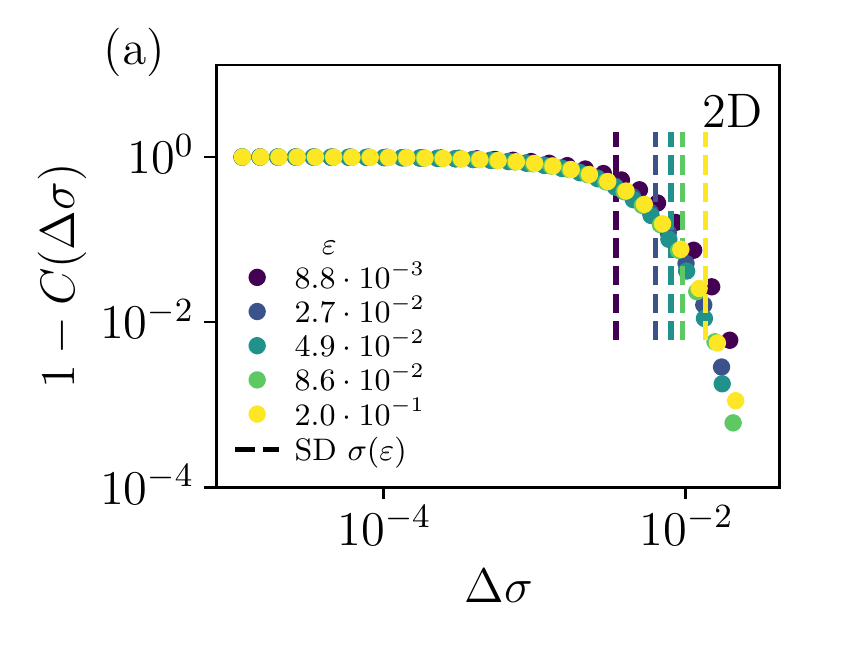}
    \includegraphics[width=0.4\linewidth]{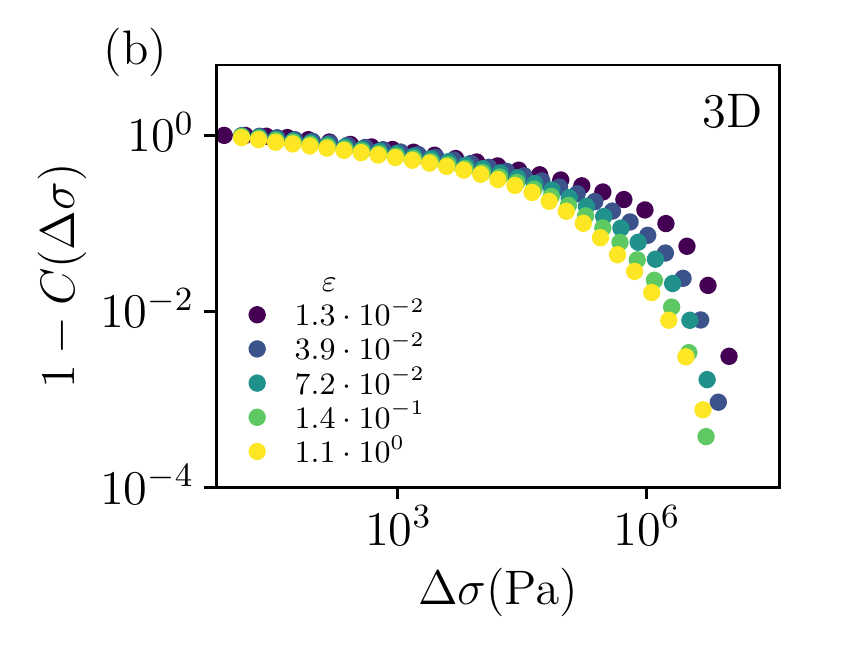}
    \caption{ Strain-dependent complementary cumulative distributions of $\Delta \sigma$ in (a) 2D and (b) 3D systems for avalanches starting with varying strain level (color coding). In (a), the dashed lines illustrate the standard deviation of stress-strain curves with the corresponding strain.   }
    \label{fig:stressincrement}
\end{figure*}

The eventual decrease of $\frac{N(\sigma - \langle \sigma \rangle = 0)}{\mathrm{d}\varepsilon}$ at large strains has a connection to the 'yield stress', i.e. the system-specific stress required to enter plastic flow. 
One reason for the dropping rate is the less and less frequent avalanches but additionally, the stress increments between the avalanches get smaller and the standard deviation from the average stress-strain curve increases compared to $\langle \Delta \sigma \rangle$  as seen in Fig. \ref{fig:stressincrement} which shows the strain-dependent complementary cumulative distributions of $\Delta \sigma$.
Therefore, the single systems 'freeze' their relative behavior with respect to $\langle \sigma \rangle$ and decide between being a "stronger" or a "weaker" sample.
This then affects the predictability of single system stress-strain curves interestingly: 
Fig. \ref{fig:intersectionrate}a illustrates also the goodness of a neural network (NN) fit of the 2D initial dislocation structures to the   ensuing stress-strain curves computed in Ref.~\cite{salmenjoki2018machine}. 
The NN fit is best at small strains and large strains while at intermediate strains, where $\frac{N(\sigma - \langle \sigma \rangle = 0)}{\mathrm{d}\varepsilon}$ is high and most of the avalanche activity happens, the score of the fit drops to a clear minimum. 
Thus, predicting the stress-strain curve accurately is impossible at strains, where the systems oscillate below and above the average behaviour.

\subsection{Avalanche correlations in pre-strained and precipitate-dominated systems}

What then happens to these correlations, that push the system response towards the average, if the system has some deformation history or includes disorder in the form of quenched pinning points?
To study the effect of deformation history, we consider 5000 2D DDD systems that have been first pre-strained up to a strain $\varepsilon_{ID} = 0.2$ and then relaxed to a new, meta-stable initial state \cite{salmenjoki2018machine}. 
Avalanches that then occur during a quasistatic stress ramp in these pre-strained systems are analyzed with the same scheme as applied above on the basic systems.

\begin{figure*}[h]
    \centering
    \includegraphics[width=0.3\linewidth]{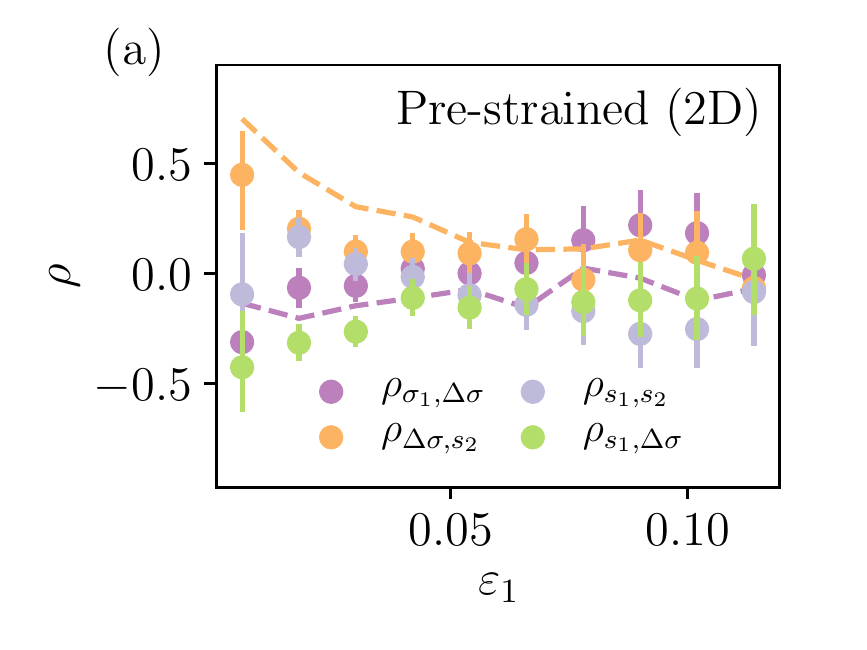}
    \includegraphics[width=0.3\linewidth]{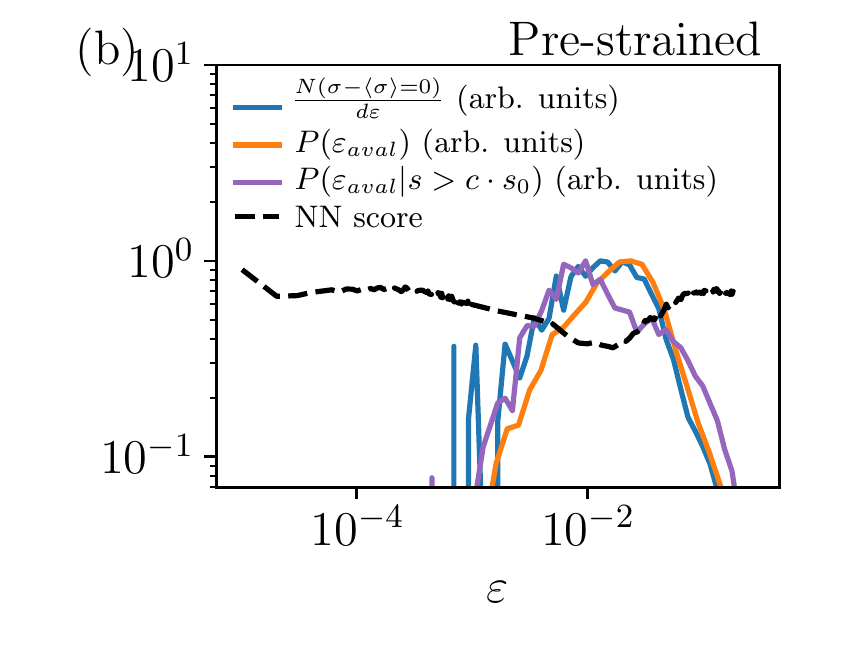}
    \includegraphics[width=0.3\linewidth]{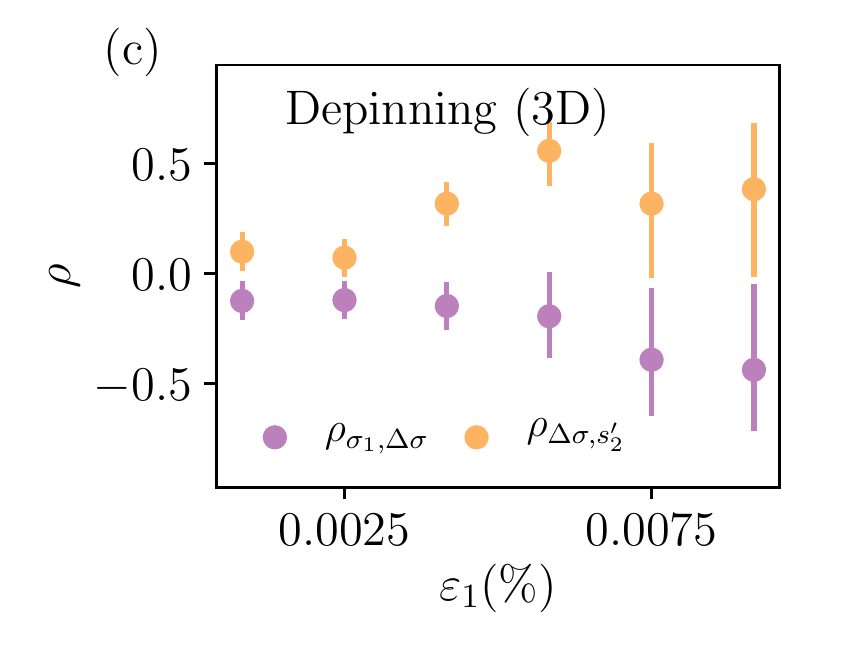}
    \caption{ (a) Spearman correlation coefficient between subsequent avalanche parameters in pre-strained 2D systems along the stress-strain curve. For reference, the color-coded dashed lines show $\rho_{\sigma_1, \Delta \sigma}$ and $\rho_{\Delta \sigma, s_2}$ measured in the systems without pre-strain. (b) The rate of intersections with the average stress-strain curve and distribution of avalanche starting strains and NN score for pre-strained systems. (c) Spearman correlation coefficient between subsequent avalanche parameters in 3D systems with precipitates (depinning). The used large avalanche thresholds are (a) $c=0.15$ and (b) $c=0.2$ }
    \label{fig:prestrain_depinning}
\end{figure*}

Fig. \ref{fig:prestrain_depinning}a shows the Spearman correlation coefficients  between relevant parameters of two subsequent avalanches in pre-strained systems. 
Generally, the correlations are similar compared to the 2D case: $\rho_{\sigma_1, \Delta \sigma}$ is negative for small strains and $\rho_{\Delta \sigma, s_2}$ is positive, both steering the single system response towards the average. 
But there are two notable differences too: 
First, the correlations seem to decline already with smaller strains than in the 2D case.
Although this is not that clear in the rate of intersections with the average curve in Fig. \ref{fig:prestrain_depinning}b which resembles that of Fig. \ref{fig:intersectionrate}a, the predictability in the form of the score of the NN fit increases more sharply in pre-strained systems. 
Second, correlation between first avalanche size and the following stress increment, which was negligible in the basic 2D case, is significant in pre-strained systems.

The final set of avalanches is collected from 100 3D DDD simulations where now the dislocations are accompanied by precipitates. 
Moreover, the precipitate strength is chosen to ensure the dislocation dynamics is governed by dislocation depinning with a distinct critical stress $\sigma_c = 4.4\cdot 10^{7}\,\mathrm{Pa}$ of dislocation flow \cite{salmenjoki2020plastic}. 
Opposed to pure systems following glassy dislocation dynamics, pinning force of disorder becomes the dominating interaction and it changes the subsequent avalanche correlations entirely as is seen in Fig. \ref{fig:prestrain_depinning}c.
With precipitates,  the correlations are insignificant at small strains but start to increase at strains close to the plateau of the stress-strain curve, although at larger strains the number of avalanches decreases causing larger error estimates on the magnitude of the correlation. 
Thus at small strains, the avalanches occur randomly as dislocation segments break away from pinning points.
At large strains, the (possible) correlations follow from the approaching $\sigma_c$: $\rho_{\sigma_1,\Delta \sigma} \approx -0.5$ means that, as a larger avalanche occurs with large stress, the following avalanche starts after a smaller stress increment. 
Similarly $\rho_{\Delta \sigma, s_2} \approx 0.5$ after larger stress increments between the avalanches, the following avalanche will be larger.
Therefore with depinning, the correlations between avalanches start to push single curves towards the average curve (i.e. $\sigma_c$) at large strains as opposed to 'pure' samples where the correlations were observed in the small strain region.

\subsection{Randomly generated stress-strain curves}

To further elaborate on the dislocation systems' tendency to have mechanical response follow the average, we refine the idea of randomly drawing stress-strain curves as introduced in \cite{kapetanou2015statistical}.
In our simulations with quasistatic stress ramp, the resulting system response is a stress-strain curve consisting of two recurring building blocks with three parameters: there are the avalanches with sizes $s$, which are separated by (near) linear increase of stress $\Delta \sigma$ during strain increment of $\Delta \varepsilon$ (Fig.~\ref{fig:correlation}a).
Thus, the idea is to resolve the independent distributions of the three variables -- $p(s)$, $p(\Delta \sigma)$ and $p(\Delta \varepsilon)$ -- and draw samples from the distributions to mimic stress-strain curves of the simulations.

We approximate the distributions by utilizing a Markov chain Monte Carlo (MCMC) method called Metropolis-Hastings \cite{gelman2013bayesian}.
Because the distributions evolve as the simulations progress, we divide the avalanches to eight strain bins and build separate samplers for each variable in each strain bin.
We neglect all the small noise avalanches as we fit the samplers. 
To ensure the convergence of our MCMC samplers, we measure the potential scale reduction factor and compare the distributions of the drawn samples to the results of the simulations \cite{gelman2013bayesian}. 
For comparison, we also test a multivariate distribution, $p(s, \Delta \sigma, \Delta \varepsilon)$, which should capture part of the inter-avalanche correlations observed in this paper, but unfortunately this converges only for the 2D and pre-strained cases.

\begin{figure*}[htb!]
    \centering
    \includegraphics[width=1\linewidth]{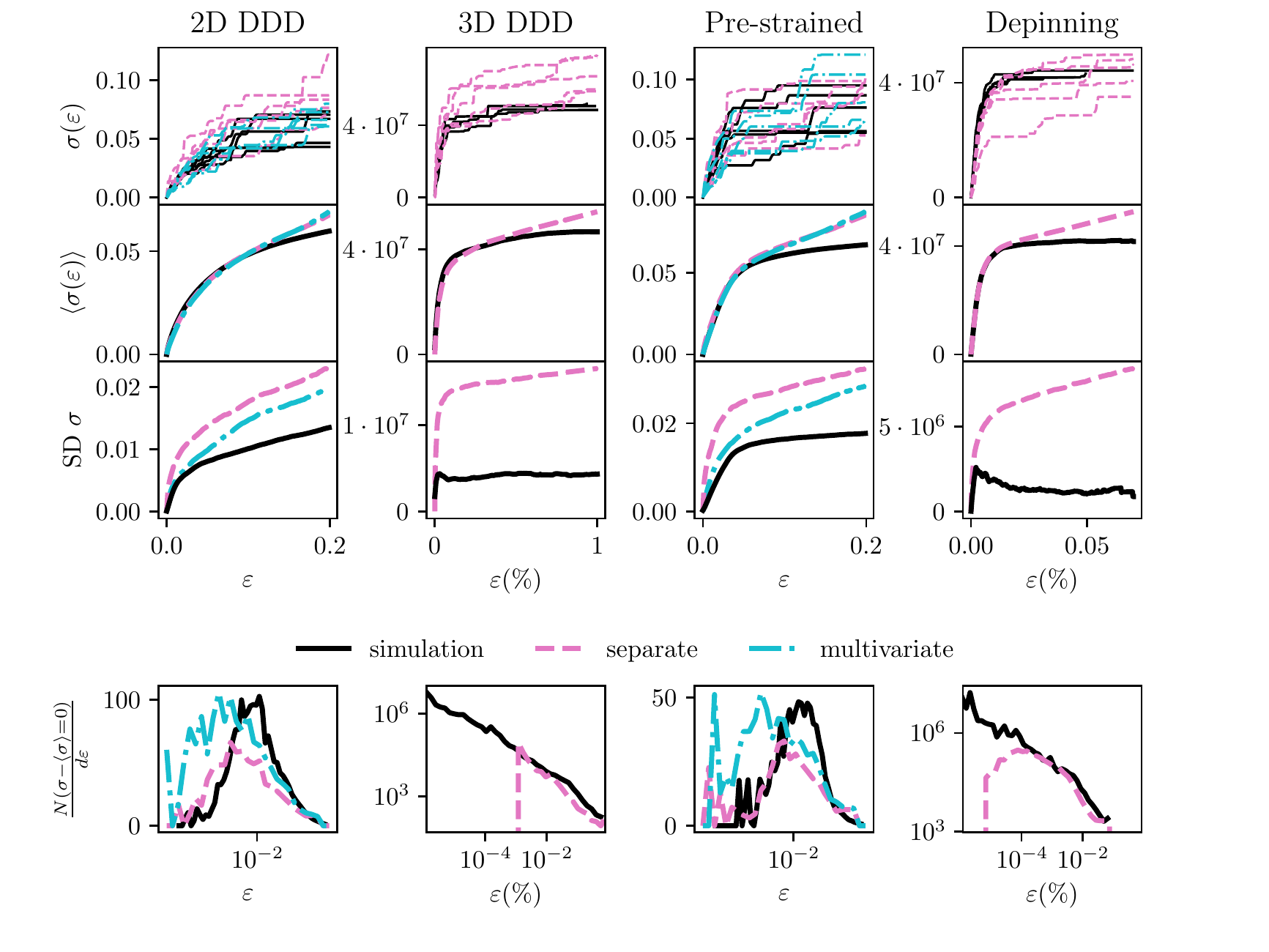}
    \caption{ MCMC sampled random stress-strain curves. Different columns from left to right correspond to different datasets of basic 2D, basic 3D, pre-strained (2D) and depinning (3D) simulations. In the top row we have examples of randomly drawn stress-strain curves (dashed lines for curves from separate distributions, dash-dotted for multivariate distributions where applicable; see text for more information) compared to simulated curves (solid lines). The second row from top shows the average stress-strain curves of 1000 randomly drawn samples along with the simulation results and the third row from the top has the standard deviation from the average curve. 
    The bottom row has the rate of intersections with the average curve.     }
    \label{fig:mcmc}
\end{figure*}

The resulting random stress-strain curves are illustrated in Fig. \ref{fig:mcmc}. 
The different rows of the figure correspond to example stress-strain curves (top row), average stress-strain curve, standard deviation from the average curve and the rate of intersecting the average curve (bottom row) for the different systems (2D, 3D, pre-strained, depinning) separated in distinct columns. 
Starting from the top row, the randomly drawn curves have the same staircase-like shape as the simulated curves as expected. 
The average shape of the randomly drawn curves also follows the simulation results nicely at least with small and intermediate strains.
However, there is a clear stress overshoot which emerges for every case, thus hinting at a possible systematic error.
But instead of MCMC sampling, the error arises already from the simulations: the avalanche data that we use to build the samplers does not include those avalanches that are unfinished as the simulation finishes. 
Therefore, large avalanches are underrepresented in the fitting data and the randomly drawn curves have larger slope at large strains than required. 

The standard deviation from the average (second row from the bottom) shows a clear distinction between the simulated and the random stress-strain curves. 
Using independent separate distributions for samplers yields curves that have a significantly wider spread at all strains than the simulated curves which, again, highlights the tendency of the dislocation systems to follow the average response. 
In 2D and pre-strained cases where we are able to draw curves also from the converged multivariate distribution $p(s, \Delta \sigma, \Delta \varepsilon)$, standard deviation is slightly closer to the simulated than the separate distributions. 
As the multivariate distribution contains some information of the correlations between $\Delta \sigma$ and $s_2$ and $\sigma_1, \Delta \sigma$ (through the different samplers along the stress-strain curve), the wider spread of the curves could imply some longer-range correlations in the avalanche time series.

Finally the bottom row shows the rate of intersections with the average curve compared to the simulations.
With 2D and pre-strained data, the multivariate random curves show more similar magnitude and shape of $\frac{N(\sigma - \langle \sigma \rangle = 0)}{\mathrm{d}\varepsilon}$ then the separate distribution random curves as expected from the smaller standard deviation. 
In 3D and depinning, the rate magnitude is quite close to the simulation results even though the curves are drawn from separate distributions. 
However, there the small strain behaviour is not captured by the sampler because the initial state is not entirely stable and many small avalanches occur in simulations with ParaDiS (as mentioned above), and our samplers are fitted without the noisy avalanche data. 

\begin{figure*}[htb]
    \centering
    \includegraphics[width=0.4\linewidth]{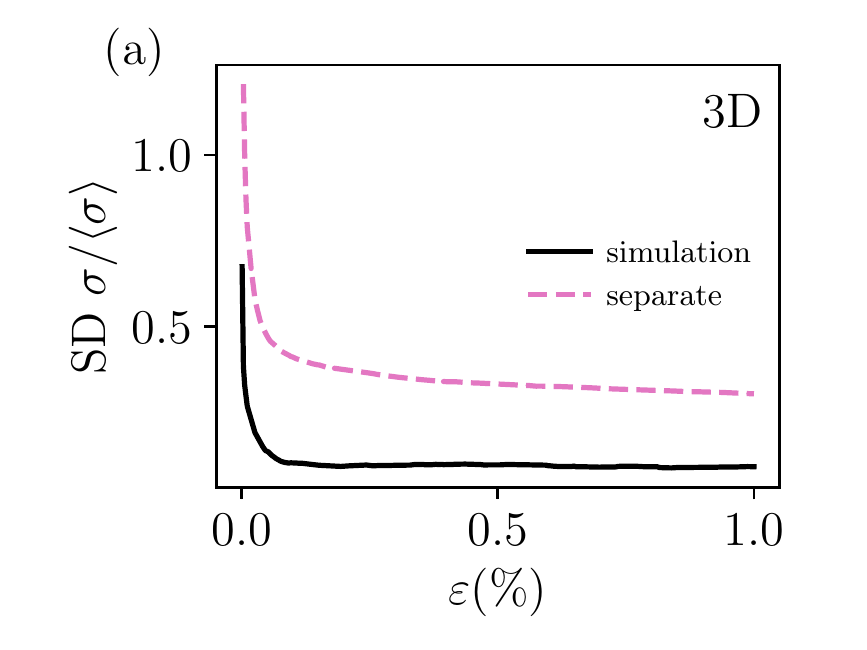}
    \includegraphics[width=0.4\linewidth]{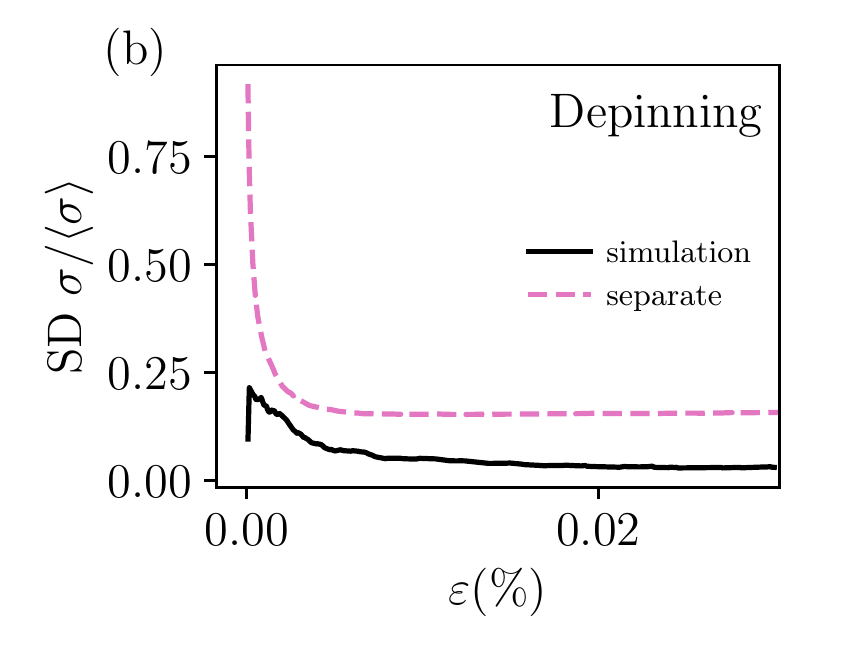}
    \caption{Standard deviation of stress values divided by the average stress as a function of strain in (a) 3D and (b) depinning simulations (solid line) compared to the case of randomly drawn curves from separate distributions (dashed line).   }
    \label{fig:ss_std}
\end{figure*}

The comparisons of reconstructed and simulated stress-strain curve sets lead us to the final question of the averaging of stress with increasing strain. This relates directly to the question of what strain values are physically justified for an effective definition of a yield stress. Figure \ref{fig:ss_std} shows for the 3D cases how the standard deviation of sample-dependent stress values decays with $\varepsilon$. We find without and with precipitates similar behaviors even though as noted the strain ranges are different. The standard deviation decays quite fast with strain. It is notable that reconstruction leads to much larger variation.

\section{Conclusions}
\label{sec:conclusions}

In this work, we have studied the coupling of avalanche dynamics in plastic deformation with the sample response. We show that the process is fundamentally different from usual avalanching systems, where the presence of a dynamical phase transition between active (plastic flow) and passive phases makes only the proximity of the critical point (here, yield stress) interesting due to the mechanism of a diverging correlation length. Instead, in the absence of pinning points (precipitates) interfering with dislocation motion the subsequent avalanches exhibit correlations, which decay along the stress-strain curve, and result in a reduced scatter of the individual stress-strain curves around the average response. The main features of our results are independent of dimension (2D vs 3D), and are present also both in pre-strained samples or dislocation systems with quenched pinning points. The case with a true depinning 
transition has qualitatively quite similar behavior to all the other cases as Fig.~\ref{fig:mcmc} shows in particular in stress-strain reconstruction, even if the correlations tend to increase as the critical stress of the depinning is approached. A signature of the depinning critical point is also flattening of the stress-strain curve as the critical stress is approached, due to divergent size of dislocation avalanches at the critical point.

These issues are best explored in the context of studies of plasticity on the level of small, micron-size samples. The classical case would be the compression of micropillars, where as noted already fluctuations and avalanches are omnipresent~\cite{uchic2004sample,dimiduk2005size,ispanovity2010submicron,papanikolaou2012quasi}. The role of big avalanches and the tendency to follow (or not follow) the average behaviour at a given strain may be investigated by gathering enough statistics and by paying attention to the correlations as we have done  here. An additional feature is that the presence of self-averaging at larger strains seems to imply that the extra disorder (precipitates here) controls the dislocation dynamics. Precipitation strengthening is thus coupled to this feature.

One should also point out the fact that in the 3D case, if one defines the yield stress to correspond, e.g., to 0.1 \% plastic strain our simulations show that the width of the yield stress distribution is significantly narrower than one might expect assuming uncorrelated avalanches (Fig.~\ref{fig:ss_std}). It would be interesting to explore such correlations and their effect on the shape of the stress-strain curve in deformation experiments, and extend our study to the case of polycrystalline samples. Finally, avalanche correlations could be looked for also in amorphous plasticity~\cite{budrikis2017universal}.


\begin{acknowledgements}
LL acknowledges the support of the Academy of Finland via the Academy Project COPLAST (project no. 322405). 
HS acknowledges the support from Finnish Foundation for Technology Promotion. 
MA acknowledges support from the European Union Horizon 2020 research and innovation
programme under grant agreement No 857470 and from European Regional Development Fund via Foundation for Polish Science International Research Agenda PLUS programme grant No MAB PLUS/2018/8.
The authors acknowledge the computational resources provided by the Aalto University 
School of Science ``Science-IT'' project, as well as those provided by CSC (Finland).
\end{acknowledgements}

%

\vspace{0.5cm}



\end{document}